\documentstyle[aps,prl,epsfig,multicol,amsmath]{revtex}

\renewcommand{\narrowtext}{\begin{multicols}{2}\global\columnwidth20.5pc}
\renewcommand{\widetext}{\end{multicols}\global\columnwidth42.5pc}

\begin{document}

\title{Coulomb Drag in Coherent Mesoscopic Systems}

\draft

\author{Niels Asger Mortensen$^{1,2}$, Karsten Flensberg$^2$, and
  Antti-Pekka Jauho$^1$}

\address{$^1$Mikroelektronik Centret, Technical University of Denmark,
  \O rsteds Plads bld. 345 east, DK-2800 Kgs. Lyngby, Denmark\\ $^2$\O
  rsted Laboratory, Niels Bohr Institute, Universitetsparken 5,
  DK-2100 Copenhagen \O, Denmark}

\date{\today}

\maketitle

\begin{abstract}
  We present a theory for Coulomb drag between two mesoscopic systems.
  Our formalism expresses the drag in terms of scattering matrices and
  wave functions, and its range of validity covers both ballistic and
  disordered systems.  The consequences can be worked out either by
  analytic means, such as the random matrix theory, or by numerical
  simulations.  We show that Coulomb drag is sensitive to localized
  states, which usual transport measurements do not probe. For chaotic
  2D--systems we find a vanishing average drag, with a nonzero
  variance. Disordered 1D--wires show a finite drag, with a large
  variance, giving rise to a possible sign change of the induced
  current.
\end{abstract}

\begin{figure}
\begin{center}
\epsfig{file=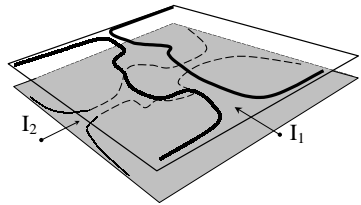, width=4.7cm,clip}
\end{center}
\caption{Schematic geometry of a mesoscopic Coulomb drag
  experiment~\protect\cite{dots}. }
\label{fig:layout}
\end{figure}

\narrowtext
Moving charges in a conductor exert a Coulomb force on the
charge-carriers in a nearby conductor, thus inducing a drag-current
(see Fig.~\ref{fig:layout}). This happens whenever the distance
between the two conductors is of the same order as the average
distance between charge carriers. In recent years Coulomb drag in
two-dimensional systems has been studied extensively~\cite{review} and
has provided valuable information about the interactions between
adjacent extended electron gases.

Coulomb drag of mesoscopic structures has been addressed in the case
of 1D--systems both within the Boltzmann equation
approach~\cite{boltzmann} and for Luttinger liquids with strong
interwire interactions~\cite{idealdrag}.

The study of fluctuations in the mesoscopic regime was recently
initiated by Narozhny and Aleiner~\cite{naro00}, and it was
established that fluctuations will dominate at temperatures smaller
than the Thouless energy. This was predicted to be the case even for
large extended samples, such as those used in the 2D
experiments~\cite{review}. While Ref.~\cite{naro00} concentrated on
structures larger than the phase-breaking length, $\ell_\phi$, here we
study Coulomb drag of mesoscopic samples smaller than $\ell_\phi$.
Experimentally there is so far only little work on drag in structures
with $L<\ell_{\phi}$~\cite{mesoexp}. We believe this would be an
extremely promising new direction for the study of mesoscopic
transport properties, since it gives an opportunity to directly study
interaction and correlation effects in mesoscopic structures.
Especially disordered mesoscopic systems are known to exhibit
interesting and unusual physics and the same can be expected for
disordered Coulomb drag systems -- perhaps even more so because
Coulomb drag in addition to the dependence on the transmission
properties also has a strong dependence on the nature of the wave
function inside the mesoscopic region. We note that Coulomb coupling
also has interest in other contexts, such as capacitive coupling of a
mesoscopic conductor to the environment, charge pumping in quantum
dots, or spin-polarized transport~\cite{coulomb}.

Before presenting the technical details we state our main results. We
develop a formalism for studying drag in mesoscopic systems, and apply
it to a number of special cases. In the case of 1D--wires, studied
numerically, we find that even a small amount of disorder induces
fluctuations, such that the drag can exceed the ballistic limit, be
strongly suppressed, or even change sign.  The sign change is a
general feature of mesoscopic drag, which we also demonstrate for
chaotic systems. Here arguments based on random matrix theory show
that the drag is zero on average, while the fluctuations are finite.
The zero average drag can thus be taken as a test of the degree of
ergodicity of the system under investigation. Furthermore, we address
the importance of localized states in the sample~\cite{localization}.
While localized states do not usually effect the ordinary transport
properties, they turn out to be important for the transconductance.
The reason is that the electron-electron interaction allows for
transitions in and out of the localized states, which become visible
at temperatures smaller than the level spacing, giving rise to peaks
in the transconductance when they cross the Fermi level. We also find
a temperature dependence which is very different than the
$T^2$--dependence found for extended states.

{\it General formulation --} Using linear response theory similar to
Refs.~\onlinecite{kame95,flen95} we find the Coulomb drag to second
order in the interaction between mesoscopic subsystems, $U_{12}$,
taking the isolated systems to be otherwise non-interacting. The
general formula for the dc transconductance in the case of two
mesoscopic conductors, as illustrated in Fig. 1, is given by
\begin{eqnarray}
G_{21}&=& \frac{e^2}{h}\int {\rm d}{\bf r}_1 {\rm d}{\bf r}_2
{\rm d}{\bf r}_1' {\rm d}{\bf r}_2'\, U_{12}({\bf
r}_1,{\bf r}_2) U_{12}({\bf r}_1',{\bf r}_2')\nonumber\\
&&\quad \times \hbar \int_{-\infty}^\infty {\rm d}\omega\,
\frac{\Delta_1(\omega,{\bf r}_1,{\bf r}_1') \Delta_2(-\omega,{\bf
r}_2,{\bf r}_2')}{2 kT\, \sinh^2(\hbar\omega/2kT)},\label{G21}
\end{eqnarray}
where $\Delta$ is the three point correlation function $\langle
\hat{I} \hat{\rho} \hat{\rho}\rangle$, as explained in
Ref.~\cite{flen95}.  Eq.~(1) generalizes the results of
Ref.~\cite{kame95,flen95} to systems with broken
translation-invariance.  For the case of mesoscopic conductors it
becomes
\begin{eqnarray}
&&\Delta_i(\omega,{\bf r},{\bf r}')=-2i\pi^2\hbar
\sum_{\beta}\theta^i_\beta({\bf r},{\bf
r}',\varepsilon_\beta-\hbar\omega)\nonumber\\
&&\quad \times \big[n_{F}(\varepsilon_{\beta}-\hbar\omega)-
n_{F}(\varepsilon_{\beta})\big] + \big({\bf r}\leftrightarrow {\bf
r'};\, \omega\rightarrow -\omega\big). \label{Deltadef}
\end{eqnarray}
Here
\begin{equation}
\theta_\beta^i({\bf r},{\bf r}',\varepsilon)=\sum_{\alpha \gamma}
I^i_{\alpha\gamma}\rho^i_{\alpha\beta}({\bf r})
\rho^i_{\beta\gamma}({\bf
r}')\delta(\xi_\alpha)
\delta(\xi_\gamma),\label{thetadef}
\end{equation}
where $\xi_{\alpha}=\varepsilon_{\alpha} -\varepsilon$ and $i$ labels
the subsystem. The matrix elements are given by
$I^i_{\alpha\gamma}=\langle \alpha|\hat{I}^i|\gamma\rangle$ and
$\rho^i_{\alpha\beta}({\bf r})=\langle \alpha|{\bf r}\rangle\langle
{\bf r}|\beta\rangle$, where $|\alpha\big>$'s are the eigenstates of
the uncoupled subsystem with energies $\varepsilon_\alpha$.  Using
scattering states as the basis we get
$I^i_{\alpha\beta}=\frac{\hbar}{2m}
\delta_{\varepsilon_\alpha,\varepsilon_\beta}j_{\alpha\beta}$, where
the matrix $j$ can be expressed in terms of the $2N\times 2N$
scattering matrix $S$~\cite{datta} as $j=\left(\tau^3-S^\dagger\tau^3
  S\right)$. Here, $\tau^3_{nn'}=\pm\delta_{nn'}$ with plus for $n$
belonging to right moving scattering states and minus for the left
moving states.

Some general features immediately follow from Eq.~(\ref{G21}).  The
usual cancellation of velocity and density of states, which is central
in the derivation of the Landauer--B\"{u}ttiker formula, occurs only
for $I^i_{\alpha\gamma}$, whereas for $\rho^i_{\alpha\beta}$ this is
not the case.  Consequently, in contrast to individual subsystem
conductances $G_{ii}$, $G_{21}$ peaks at the onset of new modes in
either of the subsystems. Secondly, we notice that the sum over
$\left|\beta\right>$ mixes both propagating and evanescent modes. This
means that apart from the transmission properties also localized
states are probed by measuring drag conductance. Finally, we notice
that the outcome of Eq.~(\ref{G21}) can have any sign, which is
directly related to lack of translation-invariance.

The low temperature limit also follows readily from Eq.~(\ref{G21}).
The factor $\sinh^{-2}$ cuts off the frequency integration and we can
expand the $\Delta$'s to lowest order in $\omega$.  This gives
$\Delta\propto \omega$ with the sum over states restricted to those at
the Fermi level ($\xi_\beta^{F}=\varepsilon_{\beta}-\varepsilon_{F}$):
\begin{equation}
\Delta_i(\omega,{\bf r},{\bf r}')=4\omega \pi^2\hbar^2{\mathrm Im}
\sum_{\beta}\theta_\beta({\bf r},{\bf
r'},\varepsilon_{F})\delta(\xi_{\beta}^ {F}).
\label{DeltalowT}
\end{equation}
We immediately see that the transconductance in this limit becomes
proportional to $T^2$, in accordance with the usual Fermi liquid
result for electron-electron scattering. Note however that the low
temperature expansion breaks down when the temperature becomes smaller
than the level spacing of the discrete, {\it i.e.}  localized states,
which we discuss in detail below. At higher temperatures the
$T^2$-behavior is replaced by a weaker temperature dependence ({\it
  e.g.} for a quasi 1D--system, $G_{21}\propto T$ for $kT> \hbar
v_{F}/L$ as considered in Ref.~\cite{boltzmann}). Here we however
concentrate on the low temperature dependence.

\begin{figure}
\begin{center}
\epsfig{file=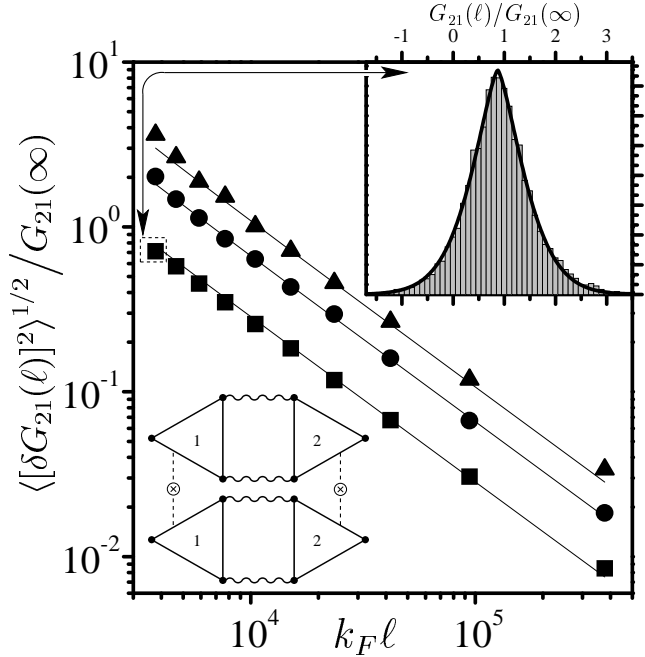, width=8.5cm,clip}
\end{center}
\caption{Relative fluctuations of the transconductance as a function
  of the mean free path for the following lengths of the two
  1D--wires: $k_FL= 100\,\pi/3$ (squares), $200\,\pi/3$ (circles), and
  $300\,\pi/3$ (triangles). The full lines are the results of
  Eq.~(\ref{1Danalytic}) which is shown diagrammatically by the lower
  left inset. The upper right inset shows a typical histogram based on
  $10^4$ disorder configurations.}
  \label{fig:1D}
\end{figure}

{\it One-dimensional wires --} Next we consider as an illustrative
example two disordered 1D--wires, which we solve both numerically and
analytically using perturbation theory. The one-dimensional case shows
that a small amount of disorder can lead to large fluctuations for the
drag response and even reverse the sign. The reason for this is that
inter-wire interaction induced forward scattering gives rise to a drag
response provided it is combined with disorder induced backscattering.
In contrast, in the case of clean wires the backscattering is induced
solely by the interwire interaction, and therefore the disordered case
is larger by a factor of order $\big<{\cal
  R}\big>U_{12}(0)/U_{12}(2k_F)$, with
$U_{12}(q)=\int_{0}^{L}\int_{0}^{L}dx_{1}\,dx_{2}\,e^{iq(x_{1}-x_{2})}
U_{12}(x_{1},x_{2})$ being the Fourier transformed interaction and
${\cal R}$ is the reflection coefficient, which is inversely
proportional to the mean free path $\big<{\cal R}\big>\simeq L/\ell$.
We can show this explicitly by considering the lowest order
perturbation theory in disorder potential, corresponding to the
diagrams shown in Fig.~2, and for long wires $k_FL\gg 1$ we find
\begin{equation}
\frac{\big<[\delta G_{21}(\ell)]^{2}\big>^{1/2}}{G_{21}(\infty)}
\simeq\frac {\big[2\big<{\cal
R}_{1}\big>\big<{\cal R}_{2}\big>U_{12}^2(2k_{F})\widetilde{U}_{12}^{2}
(0)\big]^{1/2}}{U_{12}^2(2k_{F})},\label{1Danalytic}
\end{equation}
where
\begin{eqnarray}&&\widetilde{U}_{12}^2(0)\equiv \int_{0}^{L}
    \int_{0}^{L}\int_{0}^{L}\int_{0}^{L} dx_{1}\,dx_{2}\,dx_{1}'\,dx_{2}'
   \nonumber\\
&&\quad\times U_{12}(x_{1},x_{2}) U_{12}(x_{1}',x_{2}')
\Big(1-\tfrac{2|x_{1}-x_{1}'|}{L}\Big)\Big(1-\tfrac{2|x_{2}-x_{2}'|}{L}\Big).\nonumber
\end{eqnarray}
The denominator is the result $G_{21}(\infty)\propto
U_{12}^{2}(2k_{F})$ for ballistic wires.  For the realistic case where
$U_{12}(2k_{F})\ll \widetilde{U}_{12}(0)$ we see that the fluctuations
of the drag can exceed the average value. This is in contrast to the
fluctuations of the diagonal conductance $\big<[\delta
G_{ii}]^2\big>^{1/2}$, which are vanishing compared to the mean value
$\big<G_{ii}\big>= (2e^2/h)\big(1-\big<{\cal R}_i\big>\big)\sim
2e^2/h$ in the limit of weak disorder. Fig.~2 displays the prediction
of Eq.~(\ref{1Danalytic}) along with the numerical results described
below and very good agreement is seen.

In order to solve the 1D model numerically, we study Eq.~(4) on a
lattice using the method of finite differences~\cite{datta}. The
method offers a way of studying disordered systems by ensemble
averaging over different disorder-configurations~\cite{disorder}. In
our numerical example, we use a bare long-ranged Coulomb interaction
and the Anderson model with diagonal disorder~\cite{ande58}. We have
numerically studied the drag as a function of the mean free path
$\ell$ and the length $L$, choosing the Fermi energy corresponding to
a quarter-filled band, and for a separation given by $k_{F}d=1$. We
calculate both $G_{11}$, $G_{22}$, and $G_{21}$. Since the potentials
in the two wires are uncorrelated we in general have $G_{11}\neq
G_{22}$, but $\big<G_{11}\big>\simeq\big<G_{22}\big>$ and
$\big<(\delta G_{11})^2\big>\simeq \big<(\delta G_{22})^2\big>$. Our
numerical results for distributions, mean values, and fluctuations for
$G_{ii}$ are in full agreement with the results of
Abrikosov~\cite{abri81}. In the delocalized regime $\ell \gg L$ we
find as expected that disorder has almost no effect on $G_{ii}$ and
$\big<G_{ii}\big>\sim 2e^2/h$ with very small fluctuations.
Fig.~\ref{fig:1D} shows $\big<[\delta G_{21}(\ell)]^2\big>^{1/2}$
normalized by the drag $G_{21}(\infty)$ in the ballistic regime as a
function of $k_{F}\ell$. The expected $1/\ell$-dependence is born out
by the numerical calculations and we also find that the fluctuations
increase with the length of the wires. The inset shows a typical
histogram of the drag conductance showing that depending on the
disorder configuration $G_{21}(\ell)$ can be either higher or lower
than in the ballistic regime.  Furthermore, note that in agreement
with the arguments given above the drag conductance shows a sign
reversal for some disorder realizations.

{\it Localized states --} The low temperature expansion
Eq.~(\ref{DeltalowT}), which results in a $T^2$-dependence, is only
valid if $|\beta\rangle$ belongs to a continuum of states.  To
investigate the effects due to localized states we split $\Delta$ in
two parts, $\Delta=\Delta_{d}+\Delta_{l}$, where the first term is
given by Eq.~(\ref{DeltalowT}) while the second term is due to
scattering in and out of localized states,
\begin{eqnarray}
&& \Delta _{l}(x,y;\omega ) =-2i\pi ^{2}\hbar
\sum_{\beta \in \text{localized}}
\langle x|\beta \rangle \langle \beta |y\rangle \nonumber\\
&&\times \big[\phi(x,y;\varepsilon _{\beta }
-\hbar \omega)(n_{F}(\varepsilon_{\beta }-\hbar \omega)
-n_{F}(\varepsilon _{\beta } ))  \nonumber \\
&&+\phi(y,x;\varepsilon _{\beta }+\hbar \omega)
(n_{F}(\varepsilon _{\beta }+\hbar \omega)-n_{F}(\varepsilon _{\beta } ))\big],
\end{eqnarray}
where the localized states have been chosen to be real functions, and
\begin{equation}
\phi(x,y;\varepsilon )=\sum_{\alpha \gamma }\delta \left(
\xi_{\alpha }\right) \delta \left(
\xi_{\gamma } \right) \langle \gamma
|\hat{I}|\alpha \rangle \langle \alpha |x\rangle \langle y|\gamma
\rangle.
\end{equation}
At low temperatures we can approximate $\phi(x,y;\varepsilon\pm
\hbar\omega )\approx \phi(x,y;\varepsilon_{F} )$, which allows the
temperature dependence to be extracted by integration over $\omega$ in
Eq.~(\ref{G21}). Furthermore, for temperatures less than the level
spacing the response will be dominated by the coupling to the
localized level lying closest to the Fermi level.  There are thus
three different types of contributions corresponding to the response
due to localized/delocalized states in each subsystem, $G_{21}=
\frac{e^2}{h}\left(g_{d-d}+ g_{l-d}+g_{l-l}\right)$ where
$g_{d-d}\propto T^2$.  Let us consider, say, $g_{l-d}$ in some detail.
We find
\begin{eqnarray}
g_{l-d}&\propto& \hbar\int{\rm d}\omega\,\frac{\hbar \omega\left[
n_{F}(\varepsilon_1+\hbar\omega)-n_{F}(\varepsilon_1)
\right]}{kT\sinh^2(\hbar\omega/2kT)}
\nonumber\\
&\simeq& \frac{5 kT}{\cosh[0.57\,(\varepsilon_1-\varepsilon_{\rm
F})/kT]}, \label{gl-d}
\end{eqnarray}
where $\varepsilon_1$ is the energy of the localized level lying
closest to the Fermi energy.  A similar calculation gives
\begin{equation}
g_{l-l}\propto\frac{1}
    {\cosh[(\varepsilon_1-\varepsilon_{F})/2kT]
      \cosh[(\varepsilon_2-\varepsilon_{F})/2kT]}.
\label{gl-l}
\end{equation}
The relative strengths of these terms can be estimated as
\begin{equation}
\frac{g_{d-d}}{g_{l-l}} \sim\left( \frac{kT}
{\varepsilon_{ F}}k_{F} \sqrt{\cal A} \right) ^{2}\;,\;
\frac{g_{d-d}}{g_{d-l}} \sim\left( \frac{k T}{\varepsilon_{
F}}k_{F}\sqrt{\cal A} \right) ,
\end{equation}
with $\cal A$ being the interaction area.  Thus at low temperature the
contributions due to localized states will dominate. The temperature
dependence is very different from the usual $T^2$ law, and it may even
be temperature independent if both $\varepsilon_1$ and $\varepsilon_2$
lie on the Fermi level. By adjusting the Fermi energy or system
parameters, one can use the drag response to probe the properties and
statistics of localized states.

{\it Random matrix theory --} We now discuss the statistical
properties of the transconductance. This is important in order to
determine the size of the Coulomb drag for an ensemble of disordered
mesoscopic systems, such as suggested in Fig.~\ref{fig:layout}. Our
starting point is the low temperature result (\ref{DeltalowT})
(neglecting localized states). For the calculation we need the
statistical properties of the $S$-matrix, the eigenstates, and the
eigenvalues. We assume that the region where the subsystems couple by
Coulomb interactions are disordered and that they can be described by
random matrix theory \cite{RMT}. This means that the eigenvalues and
the wave functions are assumed to be uncorrelated and furthermore that
the current matrix elements $I_{\alpha\beta}$ are uncorrelated with
the value of wave functions.  The latter follows from the fact that
the current matrix elements are independent of position and may be
evaluated outside the disordered region, and hence do not correlate
with the wave functions inside the disordered region. With these
approximations

\begin{equation}
\frac{\langle \Delta(\omega,{\bf r},{\bf
r}') \rangle}{4\omega\pi^2\hbar^2}\simeq{\mathrm Im}
\sum_{\alpha\beta\gamma}\langle
I_{\alpha\gamma}\rangle\langle\rho_{\alpha\beta}\rho_{\beta\gamma}'
\delta(\xi_\alpha^F)\delta(\xi_\beta^F)\delta(\xi_\gamma^F)\rangle.
\nonumber
\end{equation}
The average of the current matrix element is evaluated using standard
RMT~\cite{RMT}, and both with and without time reversal symmetry we
find $\langle I_{\alpha\gamma}\rangle = (\hbar/2m)\langle(\tau^3+
S^\dagger\tau^3 S)_{\alpha\gamma}\rangle \propto
\tau^3_{\alpha\gamma}$, and since the second average in $\langle
\Delta \rangle $ is symmetric with respect to interchange of $\alpha$
and $\gamma$ we get $\langle \Delta \rangle =0 $ and of course
therefore $\langle G_{21} \rangle $=0.  The fluctuations are, however,
nonzero and involve the average $\langle \Delta(\omega,{\bf r},{\bf
  r}')\Delta(\tilde{\omega},{\bf s},{\bf s}')\rangle$ and hence the
combination $\langle (S^\dagger\tau^3 S)_{\alpha\beta}(S^\dagger\tau^3
S)_{\alpha'\beta'}\rangle$, which in the limit of a large $N$ becomes
$(2N)^{-2}\delta_{\alpha\beta'}\delta_{\alpha'\beta}$.  Interestingly,
again the result is not changed by breaking of time reversal symmetry,
in contrast to the UCF case, where the results with or without an
applied $B$-field differ by a factor of 2~\cite{RMT}. The variance of
the $\Delta$ then reads
\begin{equation}\nonumber
\frac{ \langle \Delta(\omega,{\bf r},{\bf r}')\Delta(\tilde{\omega},{\bf
s},{\bf s}')\rangle}{\pi^2 \omega\tilde{\omega}} \simeq \frac{C({\bf r},{\bf
r}',{\bf s}',{\bf s})-C({\bf r},{\bf r}',{\bf s},{\bf
s}')}{(2N)^2},
\end{equation}
where $C$ is a correlation function involving four density
matrices
\begin{eqnarray} &&C({\bf r},{\bf r}',{\bf s},{\bf
s}')=\sum_{\alpha\alpha'\beta\beta'}
\big\langle\rho_{\alpha\beta}({\bf r})\rho_{\beta\alpha'}({\bf
r}')\rho_{\alpha'\beta'}({\bf s})\rho_{\beta'\beta}({\bf s}')
\nonumber\\
&&\phantom{C({\bf r},{\bf r}',{\bf s},{\bf
s}')=\sum_{\alpha\alpha'\beta\beta'}} \times
\delta(\xi_\alpha^F)\delta(\xi_\beta^F)
\delta(\xi_{\alpha'}^F)\delta(\xi_{\beta'}^F)\big\rangle \nonumber\\
&&\simeq \frac{1}{(2\pi)^4}\langle A({\bf r},{\bf r}')\rangle
\langle A({\bf s},{\bf s}')\rangle\langle A({\bf r},{\bf
s}')\rangle \langle A({\bf r}',{\bf s})\rangle
\nonumber
\end{eqnarray}
to lowest order in $1/k_F\ell$. Using the average spectral function
relevant to the 2D case $\langle A(r)\rangle\simeq
(m/2\hbar^2)\exp(-r/2\ell)J_0(k_{F} r)$, and assuming, in addition to
$k_F\ell \gg 1$ also that $\ell \gg r_s$, where $r_s$ is the screening
length, we obtain the estimate
\begin{equation}
\langle \delta G^2_{21} \rangle^{1/2}\approx 10^{-4}\cdot
\frac{e^{2}}{h}\left( \frac{kT}{\varepsilon_{F}}
\cdot \frac{U_{12}(d)}{\varepsilon_{F}}\right)^{2}
\frac{r_s^2k_{F}\sqrt{\cal A}}{\ell^2N^2}.\label{G21RMT}
\end{equation}
With typical numbers for GaAs 2DEG structures the fluctuations of the
transresistance are of the order of $0.1$ Ohm, which should be
measurable.

In conclusion, we have studied drag in mesoscopic systems and argue
that measurement of transconductance provides an interesting new
method for investigation of the electronic properties of these
systems.

We thank C.~W.~J. Beenakker and M. Brandbyge for useful discussions.

\widetext

\end{document}